\documentclass[aip,graphicx]{revtex4-1}

\usepackage{graphicx}
\usepackage{dcolumn}
\usepackage{bm}
\usepackage[usenames]{color}

\usepackage{ulem}

\draft 

\begin{document}

\title{Efficient acceleration of high-charge quasi-monoenergetic electron bunches in the blow-out regime using a few TW mid-infrared laser pulses} 

\author{Eisuke Miura}
\email{e-miura@aist.go.jp}
\affiliation{
National Institute of Advanced Industrial Science and Technology (AIST), 
Tsukuba Central 2, 1-1-1 Umezono, Tsukuba, Ibaraki 305-8568, Japan
}
\author{Shin-ichi Masuda}
\affiliation{Japan Synchrotron Radiation Research Institute (JASRI), 
1-1-1 Kouto, Sayo-cho, Sayo-gun, Hyogo 679-5198, Japan
}
\author{Eiji J. Takahashi}
\affiliation{
RIKEN Center for Advanced Photonics, 
2-1 Hirosawa, Wako, Saitama 351-0198, Japan
}

\date{\today}

\begin{abstract}
Efficient production of high-charge electron bunches in laser wakefield acceleration using a mid-infrared (MIR) laser pulse is investigated by two-dimensional particle-in-cell simulations.
Only a 2.5 TW (100 mJ, 40 fs) MIR laser pulse with a wavelength of $1.5~\mu{\rm m}$ can produce a quasi-monoenergetic electron (QME) bunch with a peak energy of 65 MeV in the blow-out regime.
The simulation results are compared with those for the experiment showing the generation of 40-60 MeV QME bunches with more than $10^8$ electrons using an 8 TW laser pulse with a wavelength of $0.8~\mu{\rm m}$.
The number of electrons in the QME bunch produced by a $1.5~\mu{\rm m}$ laser is 10 times higher than that produced by a $0.8~\mu{\rm m}$ laser, and is equivalent to $10^9$. 
The conversion efficiency from the laser energy to the QEM bunch energy is more than 10\%.
Laser wakefield acceleration by a TW-class MIR laser pulse opens the door to generating a high-charge QME bunch with high efficiency.

\end{abstract}


\maketitle 

\section{Introduction}
Laser wakefield acceleration (LWFA)~\cite{Tajima} is a promising approach to realize a compact next-generation particle accelerator using an acceleration field of more than 100 GV/m, which is a thousand times higher than that of radio-frequency accelerators. 
Research and development concerning this approach have been intensively conducted in the last two decades. 
Since the generation of quasi-monoenergetic electron (QME) bunches with a narrow energy spread was demonstrated in 2004~\cite{Mangles1, Geddes, Faure1, Miura1}, remarkable progress in the research has been seen.
The generation of QME bunches with energy of over 1 GeV has been demonstrated from a few-cm-long plasmas~\cite{Leemans2006, Wang2013, Kim2013, Leemans2014}, and the peak energy has been increased up to 8 GeV~\cite{Gonsalves2019}.
In addition, a femtosecond electron bunch can be produced in LWFA, because the electron bunch is accelerated by a plasma wave with a wavelength of the order of 10~$\mu{\rm m}$. 
By using this advantage, the applications of LWFA have also been explored, including the development of ultrashort high-brightness X-ray and $\gamma$-ray sources.
Recently, the development of X-ray and $\gamma$-ray sources using laser-accelerated electron bunches has been intensively conducted using various schemes such as betatron radiation~\cite{Rousse2004}, synchrotron radiation~\cite{Schilenvogit2008, Fuchs2009}, and inverse Thomson scattering~\cite{Schwoerer2006, Phuoc2012, Chen2013, Miura2014}.
To realize high-brightness X-ray sources, one of the key issues is the production of high-charge electron bunches.
For efficient acceleration of high-charge electron bunches, the control of electron injection and trapping into a plasma wave is crucially important, and various schemes such as colliding pulse~\cite{Faure2004}, shock injection~\cite{Buck2013}, and ionization injection~\cite{Pak2010,McGuffey2010} have been also investigated. 

The ponderomotive force of an intense laser pulse is the driving force for exciting the plasma wave, which forms the accelerating electric field.
The ponderomotive force $f_p$ is given by $f_p=-e^2/(4m_e\omega_L^2)\nabla E_L^2=-m_ec^2/4\nabla a_0^2$.
Here, $e$, $m_e$, $c$, $E_L$, $\omega _L$, and $a_0$ are the electron charge, electron mass, speed of light in a vacuum, laser electric field, laser frequency, and normalized vector potential, respectively.
The normalized vector potential is given by $a_0=eE_L/(m_ec\omega _L$).
The ponderomotive force is proportional to the square of the laser wavelength.
With a longer wavelength laser pulse in the mid-infrared (MIR) region, it is possible to efficiently excite a large amplitude plasma wave and produce high-energy electron bunches with high charge.
In almost all reported recent experiments, however, Ti:sapphire laser systems delivering a laser pulse with a wavelength of 0.8~$\mu{\rm m}$ and a duration from 10 to 100 fs have been used.
This is because only the Ti:sapphire laser technologies can routinely create pulses with few-tens femtosecond duration, realizing
relativistic laser intensity ($a_0 > 1$) by its multi-TW peak power.
Therefore, the laser wavelength has not been treated as an optimization parameter for LWFA experiments, even though the wavelength of the driver laser becomes an important parameter controlling efficient electron acceleration conditions.

Currently, there are very few discussions on the dependence on a driver laser  wavelength in LWFA.
In the early stage of the research, Nd:glass lasers with a wavelength of 1.05~$\mu{\rm m}$ were used~\cite{Modena, Nakajima, Umstadter}.
This wavelength is close to that of a Ti:sapphire laser.
In addition, the electron acceleration mechanism is different due to the long pulse duration close to 1 ps.
Then, the wavelength dependence was not discussed.
Recently, Woodbury et al. have experimentally demonstrated electron acceleration in near and above critical density plasmas using a sub-TW MIR laser pulse with a duration of 100 fs~\cite{Woodbury}.
A significant increase in the charge of accelerated electrons has been observed, and efficient electron acceleration using a MIR laser pulse is suggested.
However, it is difficult to produce a high-quality QME bunch, because the acceleration mechanism will be a self-modulated laser wakefield acceleration~\cite{Modena, Nakajima, Umstadter} due to the high electron density of the plasma.
On the other hand, Zhang et al. have reported the simulation results of LWFA using an MIR laser pulse with a peak power more than several tens of TW~\cite{Zhang}.
They have shown an increase in the charge of electron bunches using a longer wavelength laser pulse.
However, the laser power and spot diameter are fixed for each wavelength and the normalized vector potential is higher for the longer wavelength laser.
Their simulation result implies that a higher intensity laser produces electron bunches with higher energy and charge.
Although their result is quite reasonable, they do not emphasize the advantage of an MIR laser pulse in LWFA.

In this work, we have conducted two-dimensional (2D) particle-in-cell (PIC) simulations under the same normalized vector potential condition to clearly understand the advantage of LWFA using an MIR pulse.
To more quantitatively discuss the simulation results, such as the laser parameters in our simulation, the experimentally derived laser parameters are employed.
For the MIR laser pulse, we suppose a multi-TW (100 mJ, 40 fs) MIR femtosecond laser system at a wavelength of 1.5~$\mu{\rm m}$~\cite{Fu2} using dual-chirped optical parametric amplification~\cite{Fu1}.
Using our developed MIR laser system, the focused intensity with the normalized vector potential $a_0>1$ will be achieved.
From the simulation results, it is shown that electron acceleration in the blow-out regime~\cite{Pukhov} is available and a QME bunch with a peak energy of 65 MeV is produced using an MIR laser pulse with a peak power of only 2.5 TW.
In addition, the pulse duration is also close to that of the present Ti:sapphire laser used in LWFA experiments.
The simulation results can be compared with our previous work~\cite{Masuda2008, Masuda2009, Miura2009} showing the generation of 40-60 MeV QME bunches with more than $10^8$ electrons using an 8 TW laser pulse with a wavelength of $0.8~\mu{\rm m}$.
The number of electrons in the QME bunch using the $1.5~\mu{\rm m}$ laser was 10 times larger than that of the $0.8~\mu{\rm m}$ laser.
Our simulation results indicate the possibility of efficient acceleration of a high-charge QME bunch using an MIR laser pulse.

\section{Results}
\subsection{Wavelength dependence}
We conducted PIC simulations for 1.5 and 0.8 $\mu{\rm m}$ lasers to discuss the dependence on a driver laser wavelength using the conditions presented in Table 1.
As a reference case using a 0.8 $\mu{\rm m}$ laser, we select our previous experiments showing the generation of 40-60 MeV QME bunches with more than $10^8$ electrons~\cite{Masuda2008, Masuda2009, Miura2009}.
The details of the simulation setup are presented in the Methods.

Figure 1 presents the simulation results for the 1.5 $\mu{\rm m}$ laser.
Rows (i) and (ii) in Fig. 1 show a snapshot of the charge density distribution in the $x-y$ plane and the density profile on the laser propagation axis, respectively.
The density distribution is shown as the charge density distribution, because macro particles contain electrons and helium ions in our simulation code~\cite{Masuda2010}.
The charge density is normalized by the ion density.
Rows (iii) and (iv) in Fig. 1 show a snapshot of the laser electric field $E_y$ distribution in the $x-y$ plane and the profile on the laser propagation axis, respectively.
The electric field is normalized as $eE_y/(m_ec\omega _L$).
The leading edge positions of the laser pulse are around the right edge of the each figure.
As shown in Fig. 1(a)(i) and (a)(ii), around $x=80~\mu{\rm m}$, a clear bubble structure is formed behind the laser pulse, 
and the self-injection of electrons occurs at this position. 
Around $x=470~\mu{\rm m}$, the trapped and accelerated electrons reach the maximum energy and the electrons form a bunch. 
In contrast, the laser pulse is depleted at this position. 

Figure 2 shows the simulation results for the 0.8 $\mu{\rm m}$ laser.
The data are the same as in Fig. 1 and are shown in rows (i) to (iv).
In columns (a) and (b) in Fig. 2, the leading edge positions of the laser pulse are around $x=190$ and $440~\mu{\rm m}$, respectively.
Around $x=190~\mu{\rm m}$, the self-injection of electrons occurs.
Although the nonlinear plasma wave is excited, the amplitude is not so high compared with the case of the $1.5~\mu{\rm m}$ laser.
Around $x=440~\mu{\rm m}$, the accelerated electrons reach the maximum energy.
However, the number of accelerated electrons is less than that for the case of the 1.5 $\mu{\rm m}$ laser. 

Figures 3(a) and (c) show the electron energy spectra as a function of the position of the laser pulse for the 1.5 and 0.8 $\mu{\rm m}$ lasers, respectively.
The spatial evolution is obtained from the electron energy spectra calculated for the electrons with the momentum angle 
with respect to the laser propagation direction in the range of $\pm 50\,{\rm mrad}$ in the moving window for every $255$ time steps. 
In both cases, arch-like trajectories with narrow widths indicate the evolution of electron energy of QME bunches accelerated and decelerated in the first bucket.
Figure 4 shows the electron energy spectra at the position where the energy reaches the maximum for each case.
Solid and dashed lines are the energy spectra for the 1.5 and 0.8 $\mu{\rm m}$ lasers, respectively. 
The energy spectrum for 0.8 $\mu{\rm m}$ laser is enhanced by a factor of 10 to make it easy to see.
The peak energies are 60 and 65 MeV, and are comparable in both cases.
In contrast, the number of electrons for the case of the 1.5 $\mu{\rm m}$ laser is approximately 10 times higher than that of the 0.8 $\mu{\rm m}$ laser.
The peak energy of the QME bunch ($\sim60~{\rm MeV}$) for the case of the 0.8 $\mu{\rm m}$ laser is in good agreement with that of our previous experimental results~\cite{Masuda2008, Masuda2009,Miura2009}. 
In the experiments of the reference case using a 0.8 $\mu{\rm m}$ laser, QME bunches of 40-60 MeV with more than $10^8$ electrons have been obtained. 
On the basis of these results, it is suggested that a QME bunch with more than $10^9$ electrons can be produced using an MIR laser pulse with a peak power of only 2.5 TW.
The estimated energy conversion efficiency from the laser energy to QME beam energy is estimated as 10\%, which is comparable to that reported in the simulation results of the blow-out regime~\cite{Pukhov,Geissler2006}.
Efficient acceleration of a high-charge QME bunch using a few TW MIR laser pulse is shown.
Although it has been reported that a longer wavelength laser can enhance the number of accelerated electrons~\cite{Zhang}, in their simulations, the laser power and spot diameter were fixed for each wavelength and the normalized vector potential was higher for the longer wavelength laser.
In contrast, our simulation results conducted under the same normalized vector potential give the obvious different features of the dependence on a driver laser wavelength  in LWFA.

Figures 3(b) and (d) show the spatial evolution of the transverse envelopes of the laser electric field, $E_y$, on the $x$-$y$ plane 
for 1.5 and 0.8 $\mu{\rm m}$ lasers, respectively.
The spatial evolutions were obtained from the transverse envelopes of the peak amplitude of the laser electric field for every $255$ time steps. 
The electric fields are normalized as $eE_y/(m_ec\omega _L$).
In the 1.5 $\mu{\rm m}$ laser, the laser pulse focused at $x=-300~\mu{\rm m}$ propagates by maintaining a high electric field
above the Rayleigh length ($\sim140~\mu{\rm m}$).
When the pulse length is shorter than the plasma wavelength, the self-guiding of a laser pulse occurs by etching the leading edge before 
diffraction~\cite{Decker}.
The etching of the leading edge is seen in Fig. 1(a)(iii) and (a)(iv).
In contrast, for the 0.8 $\mu{\rm m}$ laser, although the laser pulse propagates by forming a narrow channel, part of the laser energy is scattered.
The electric field in the channel gradually decreases with the propagation, and the intensity is lower than that of the 1.5~$\mu{\rm m}$ laser.
Because the laser pulse length is longer than the plasma wavelength, the laser pulse is modulated both longitudinally and transversely; 
the side-scattered component is seen in Fig. 2(a)(iii).

\subsection{Electron density dependence in the 1.5~$\mu{\rm m}$ laser}
It has been reported that electron density is an important parameter for the generation of QME bunches~\cite{Masuda2008, Masuda2009}. 
For a 1.5 $\mu{\rm m}$ laser, electron density dependence is investigated, and the simulation conditions shown in Table 1 are identical except for electron density.
The simulation results are shown in Fig. 5.
Figures 5(a)(c) and (b)(d) show the electron energy spectra as a function of the position of the laser pulse, 
and spatial evolution of transverse envelopes of the laser electric field $E_y$ on the $x$-$y$ plane, respectively.
Figures 5(a)(b) and (c)(d) present the results for the electron densities of $5\times 10^{18}$ and $10^{19}~{\rm cm}^{-3}$, respectively.
For the lower density case ($5\times 10^{18}~{\rm cm}^{-3}$), although the laser pulse stably propagates by forming a narrow channel, the electric field is fairly low.
Then, the number of the accelerated electrons is quite low, although the QME bunch up to 80 MeV is accelerated at $x=1000~\mu{\rm m}$.
This suggests that this electron density of $5\times 10^{18}~{\rm cm}^{-3}$ is close to the threshold density for electron self-injection into a plasma wave.
In contrast, for the higher density case ($10^{19}~{\rm cm}^{-3}$), although the laser pulse propagates, maintaining the high electric field in a narrow channel, the depletion length of the laser pulse is short. 
As shown in Fig. 5(c), the QME bunch with a large number of electrons is accelerated.
However, the peak energy is limited to 50 MeV due to high electron density.
For a much higher density case ($1.6\times 10^{19}~{\rm cm}^{-3}$, the results are not shown in Fig. 5), the side-scattered component is observed due to self-modulation of the laser pulse, 
and the laser pulse does not propagate in the long distance.
Although a considerably large number of electrons are accelerated, the maximum electron energy is limited to 30 MeV and the energy spread is large.
It is important to optimize the electron density of the plasma for the generation of QME bunches.
This result is quite reasonable on the basis of the previous reports~\cite{Masuda2008, Masuda2009}.

\section{Discussion}
Several theoretical models of bubble acceleration in the blow-out regime have been proposed~\cite{Kostyukov2004,Gordienko2005,Lu}.
In Lu's model~\cite{Lu}, the condition for stable pulse propagation and bubble formation is given in equation(1);

$k_pR \sim k_pw_0=2\sqrt{a_0}$.    ~(1)

Here, $k_p$, $R$, and $w_0$ are the wavenumber of plasma wave, bubble radius, and laser spot radius, respectively.
The relationship is available for $a_0\ge2$. 
In addition, the laser pulse length should be less than the plasma wavelength, and is ideally half of the plasma wavelength. 
In the 1.5~$\mu{\rm m}$ laser, $k_pw_0$ in the left hand side of equation(1) is equal to 4.
As shown in Fig. 3(b), the normalized electric field $a_0$ from 2 to 3 is kept in the narrow channel.
The right hand side of equation(1) is estimated to be $2.9\sim3.5$.
This condition is quite close to the optimum condition for bubble acceleration.
In contrast, in the 0.8~$\mu{\rm m}$ laser, the laser spot radius is smaller, and the normalized electric field in the narrow channel is not high 
due to side-scattering of the laser pulse.
In addition, the laser pulse length is longer than the plasma wavelength.
This condition is far from the optimum condition for the bubble formation. 
For electron acceleration in a blow-out regime, to excite nonlinear plasma waves and form bubbles, the focused intensity with the normalized vector potential $a_0\gg1$ is an indispensable condition.
Because $a_0$ is proportional to the laser wavelength, the laser intensity to obtain the fixed $a_0$ can be lower for a longer wavelength laser such as the MIR laser, that is, the large focal spot size is available.
The matched condition for bubble acceleration is achieved by the large focal spot size for the $1.5~\mu{\rm m}$ laser.

As seen in Fig. 1(i)(a), the electron sheath forms the boundary of the bubble, and the density peaks are located at the front of the laser pulse and the rear vertex of the bubble.
These peaks are formed by electrons with relativistic velocities.
Particularly, electrons around the rear vertex of the bubble are the source of electrons, which can be trapped and accelerated in the bubble.
The number of accelerated electrons is proportional to the product of the electron density and width of the sheath~\cite{Kostyukov2004}.
As seen in Figs. 1(i)(a) and 2(i)(a), the sheath width of the rear vertex of the first bucket for the $1.5~\mu{\rm m}$ laser is much wider than that of the $0.8~\mu{\rm m}$ laser.
The ratio of the product of the electron density and sheath width for the $1.5~\mu{\rm m}$ laser to that of the $0.8~\mu{\rm m}$ laser was roughly estimated to be 9.
The swallow-tail-shaped density structure of the rear vertex seen in Fig. 1(i)(a) shows that the electron motion near the rear vertex is transverse rather than longitudinal.
This electron motion leads to the transverse wavebreaking and the threshold of the wavebreaking is reduced~\cite{Bulanov1997}.
In contrast, the clear swallow-tail-shaped density structure is not seen in Fig. 2(i)(a), and it is thought that the threshold of the wavebreaking is still high.
The wavebreaking and self-trapping of electrons occur more easily for the $1.5~\mu{\rm m}$ laser rather than the $0.8~\mu{\rm m}$ laser, which will also lead to the enhancement of accelerated electrons.
As seen in Figs. 3 (a) and (c), the distance from the laser focal point to the point of electron self-injection in the 1.5~$\mu{\rm m}$ laser ($\sim400~\mu{\rm m}$) is shorter than that in the 0.8~$\mu{\rm m}$ laser ($\sim500~\mu{\rm m}$).
These considerations largely explain the enhancement in the number of accelerated electrons for the $1.5~\mu{\rm m}$ laser.
Lu's model for bubble acceleration gives the scaling related to the number of accelerated electrons~\cite{Lu}.
The number of accelerated electrons is proportional to the laser wavelength.
Our simulation result is consistent with this scaling.

In summary, it is shown that a QME bunch with a peak energy of 65 MeV is produced using an MIR (1.5~$\mu{\rm m}$) laser pulse of only 2.5 TW in the blow-out regime with 2D PIC simulations.
The number of electrons of the QME bunch is estimated to be more than $10^9$; the corresponding energy conversion efficiency from the laser energy to the QEM bunch energy is more than 10\%.
By comparing with the case of the 0.8 $\mu$m Ti:sapphire laser, the use of a longer driving laser wavelength can improve the conversion efficiency by a factor of 10.
We believe that the presented method paves the way for the efficient generation of high-charge QME bunches from the compact ultrafast MIR laser system.

\section{Methods}
The simulations are performed in a local moving window with $2000\times1000$ cells in the $x$-$y$ plane. 
The cell size and a time step are $\sim 63\times 127\,{\rm nm}^2$ and $\sim 0.17\,{\rm fs}$, respectively. 
The laser propagation direction and the polarization direction are set in the $x$- and the $y$-axes, respectively. 
The initial position of the pulse center is located at $x=-980 \mu{\rm m}$ at time $t=0$. 
The focal position in a vacuum is located at $x=-300\,\mu{\rm m}$ and $y=0\,\mu{\rm m}$. 
The helium gas density rises at $x=-920\,\mu{\rm m}$ and smoothly increases to the maximum value at $x=-350\,\mu{\rm m}$.
The gas density remains constant from $x=-350\,\mu{\rm m}$ to investigate the laser-plasma interaction for longer propagation lengths.
The profile of the leading edge of the gas jet and the focal position are set according to the condition of the reference experiments~\cite{Masuda2008, Masuda2009, Miura2009}.
The density profile is uniform in the $y$-direction. 
The PIC simulation code includes the optical field ionization process~\cite{Masuda2010}, in which the ionization probability is calculated based on the quasi-static approximation~\cite{Landau1965}. 
A macro particle representing a helium atom contains one macro particle for a helium ion and two for electrons. 
The average number of macro particles for each species is five per cell.

We conducted simulations for 1.5 and 0.8 $\mu{\rm m}$ lasers to discuss the wavelength dependence.
The simulation conditions are shown in Table 1.
The simulation parameters for the $0.8~\mu{\rm m}$ laser shown in Table 1 are set according to the experimental conditions~\cite{Masuda2008, Masuda2009, Miura2009}.
Both simulations are performed for $a_0=2$.
To obtain the same Rayleigh length, the spot radius in a vacuum for the 1.5 $\mu{\rm m}$ laser is set to be larger.
An electron density of $7.1\times 10^{18}~{\rm cm}^{-3}$ for 1.5 $\mu{\rm m}$ laser case is the optimum electron density.
 
\acknowledgments
A part of this study was supported by JSPS KAKENHI Grant Number JP16H0409, and the Matsuo Foundation 2018.
E.J.T. gratefully acknowledges financial support from Grants-in-Aid for Scientific Research Nos. 16K13704 and 17H01067, the FY 2018 Presidents Discretionary Funds of RIKEN, and MEXT Quantum Leap Flagship Program (MEXT Q-LEAP) Grant Number JPMXS0118068681.


\begin{table}[h]
\caption{Simulation conditions.}
\label{t1}
\begin{center}
\begin{tabular}{ccc}
\hline
Laser & Ti:sapphire laser & MIR laser \\
\hline
\hline
Wavelength ($\mu{\rm m}$) & 0.8 & 1.5 \\
Pulse duration (fs) & 40 & 40 \\
Laser power (TW) & 4.9 & 2.5 \\
Normalized vector potential $a_0$ & 2 & 2 \\
Spot ($1/e^2$) radius ($\mu{\rm m}$) & 5.9 & 8.1 \\
Rayleigh length ($\mu{\rm m}$) & 140 & 140 \\
Electron density (${\rm cm}^{-3}$) & $1.6\times 10^{19}$ & $7.1\times 10^{18}$ \\
\hline
\end{tabular}
\end{center}
\end{table}

\begin{figure}
\centering
\includegraphics*[width=120mm, bb=0 0 564 584]{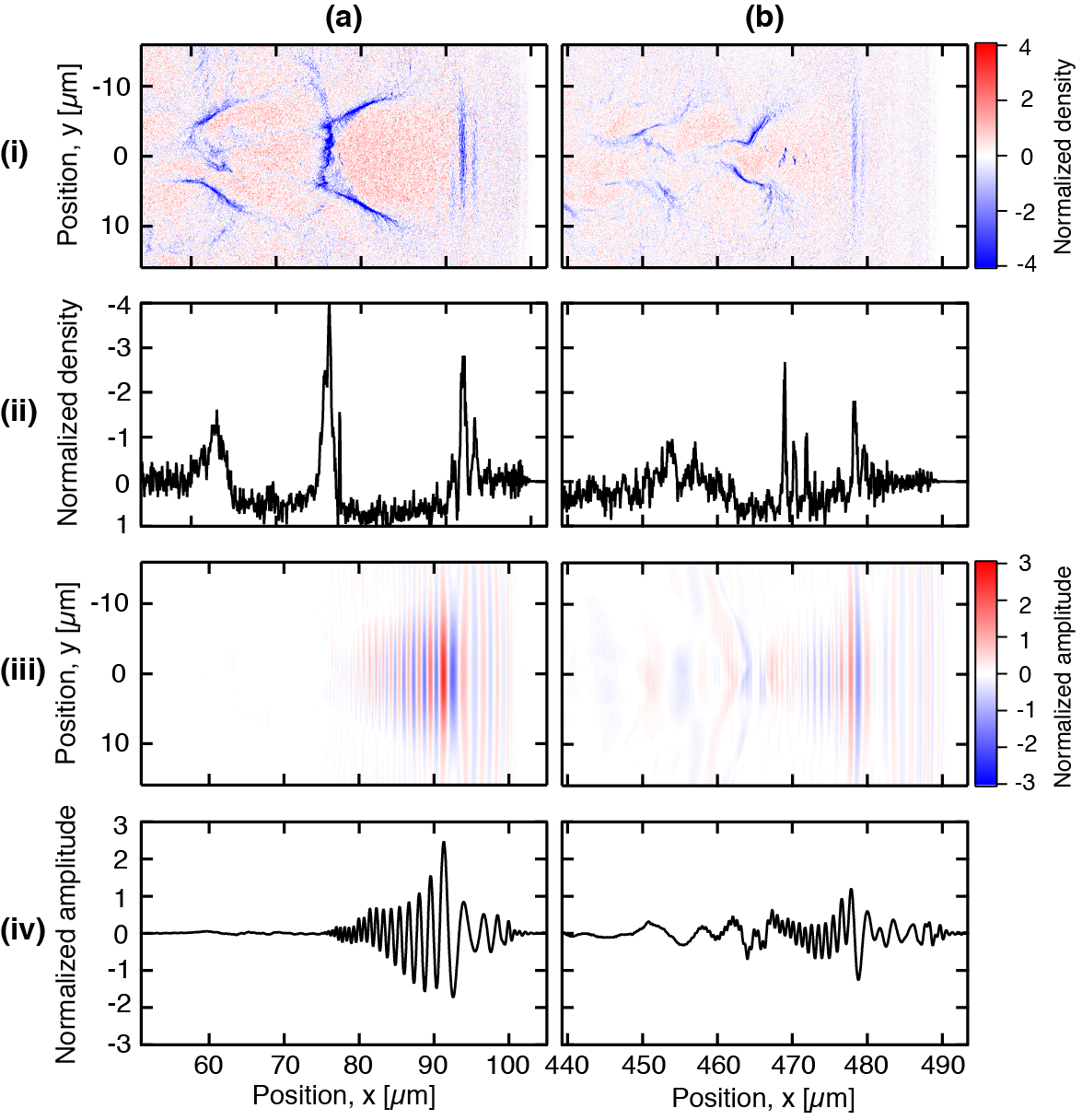}
\caption{
Simulation results for the 1.5~$\mu{\rm m}$ laser. 
Rows (i) and (ii) show a snapshot of the normalized charge density distributions in the x-y plane and the density profile on the laser propagation axis, respectively. 
Rows (iii) and (iv) show a snapshot of the laser electric field $E_y$ distribution in the $x-y$ plane and the profile on the laser propagation axis, respectively.
}
\label{fig1}
\end{figure}

\begin{figure}
\centering
\includegraphics*[width=120mm, bb=0 0 564 584]{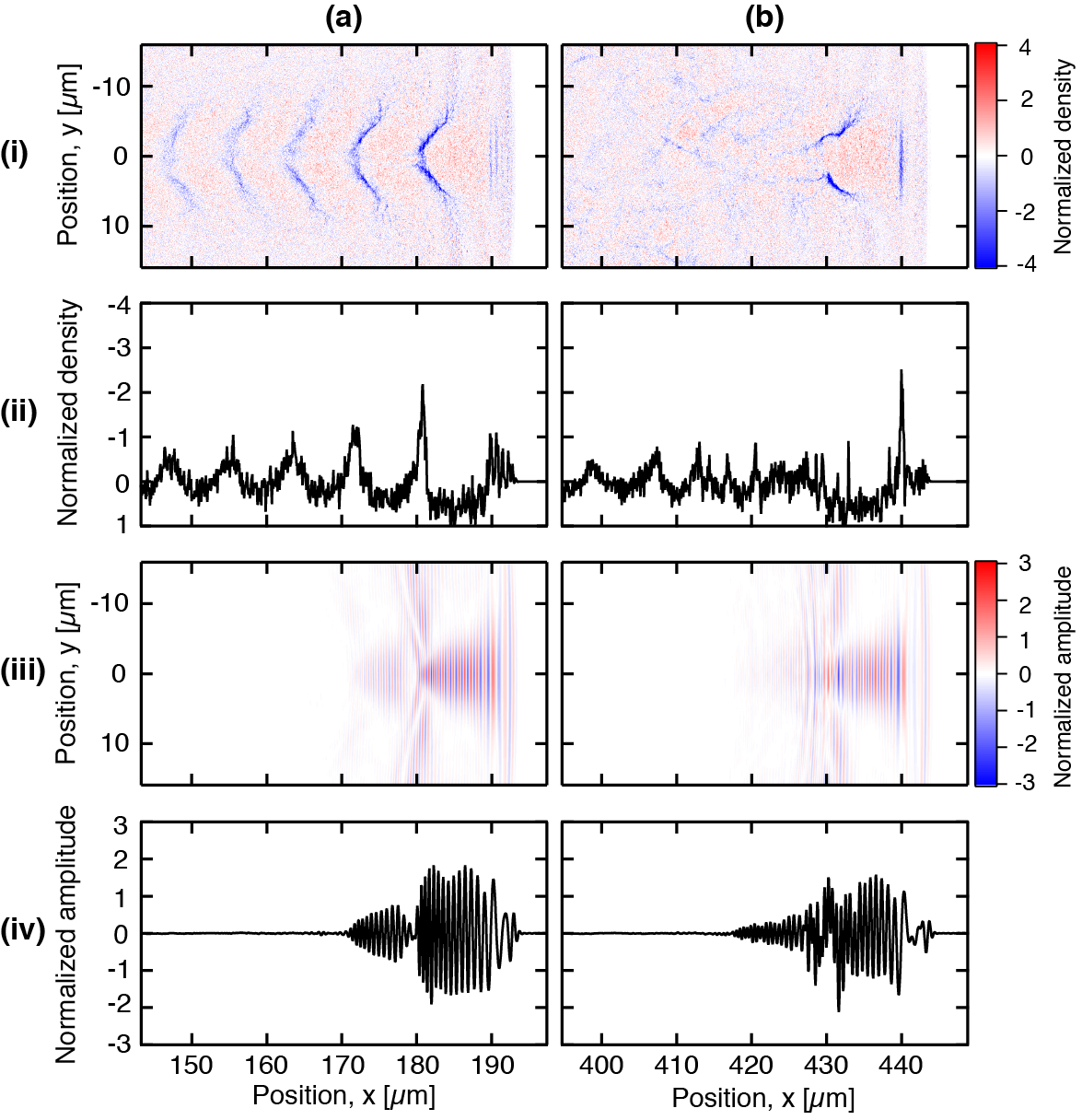}
\caption{
Simulation results for the 0.8~$\mu{\rm m}$ laser.
The data shown in rows(i) to (iv) is the same as that presented in Fig. 1.
}
\label{fig2}
\end{figure}

\begin{figure}
\centering
\includegraphics*[width=140mm, bb=0 0 576 323]{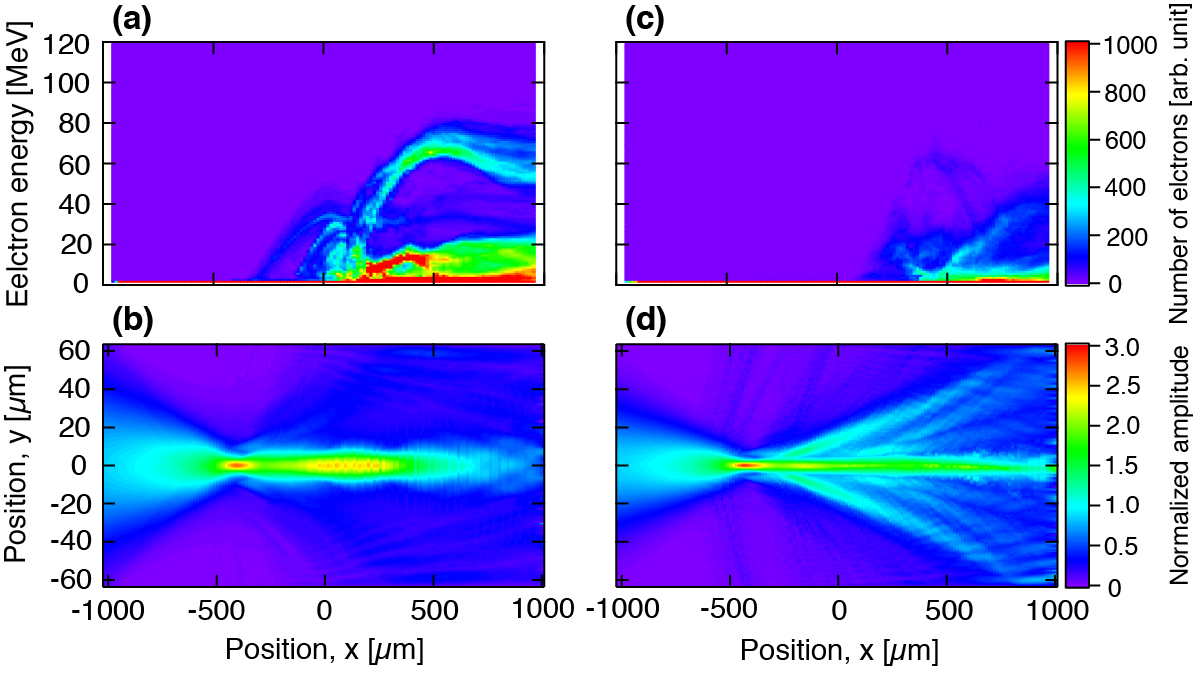}
\caption{
(a)(c) Electron energy spectra as a function of the laser pulse position.
(b)(d) Spatial evolution of transverse envelopes of the laser electric field, $E_y$, on the $x-y$ plane.
(a)(b) and (c)(d) are results for 1.5 and 0.8~$\mu{\rm m}$ lasers, respectively.
}
\label{fig3}
\end{figure}

\begin{figure}
\centering
\includegraphics*[width=80mm, bb=0 0 369 212]{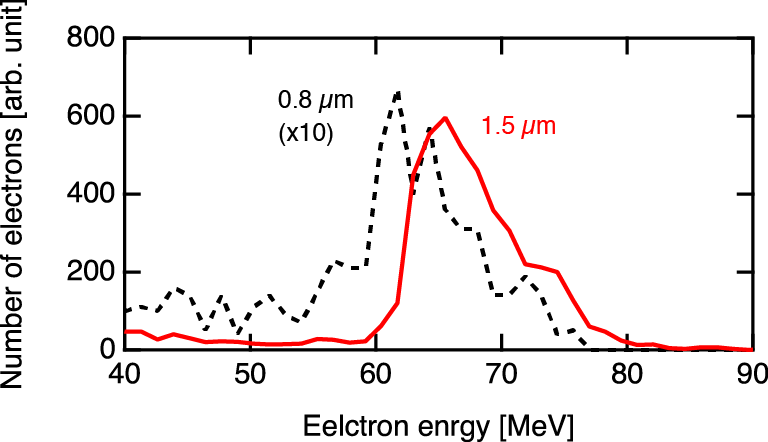}
\caption{
Electron energy spectra at the point in which the electron energy reaches the maximum.
The energy spectrum for the 0.8~$\mu{\rm m}$ laser is enhanced by a factor of 10 to make it easy to see.
}
\label{fig4}
\end{figure}

\begin{figure}
\centering
\includegraphics*[width=120mm, bb=0 0 576 323]{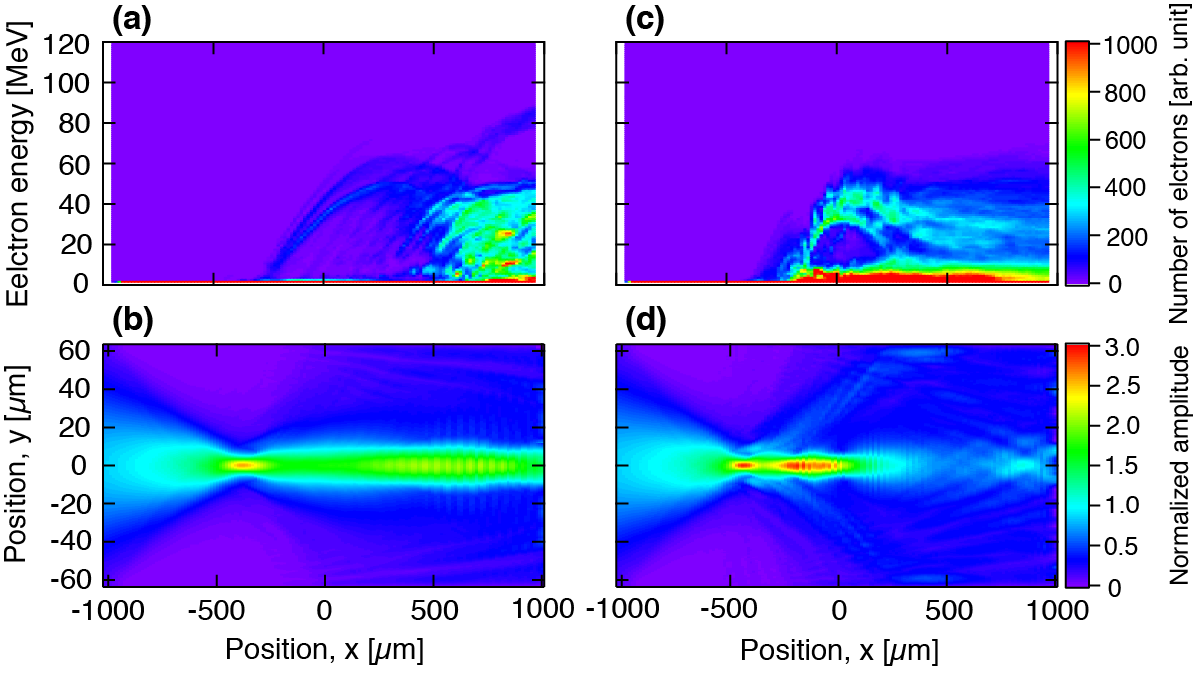}
\caption{
The results for the electron density dependence using the $1.5~\mu{\rm m}$ laser.
(a)(c) Electron energy spectra as a function of the position of the laser pulse.
(b)(d) Spatial evolution of transverse envelopes of the laser electric field, $E_y$, on the $x-y$ plane.
(a)(b) and (c)(d) are results for the electron density of $5\times 10^{18}$ and $10^{19}~{\rm cm}^{-3}$, respectively.
}
\label{fig5}
\end{figure}


\end{document}